\documentstyle[12pt,aaspp4,epsf]{article}

\received{23 January 1996}
\accepted{}

\lefthead{D. Merritt \& B. Tremblay}
\righthead{Two Families of Elliptical Galaxies}
\slugcomment{Submitted to The Astronomical Journal}

\begin{document}

\title{Evidence from Intrinsic Shapes for Two Families of Elliptical 
Galaxies}

\author{B. Tremblay \& D. Merritt}
\affil{Department of Physics and Astronomy, Rutgers University,
    New Brunswick, NJ 08855}

\begin{abstract}
Bright elliptical galaxies have a markedly different distribution
of Hubble types than faint ellipticals; the division 
occurs near $M_B=-20$ and bright ellipticals are rounder on average.
The Hubble types of galaxies in both groups are narrowly clustered, 
around E1.5 in the case of the bright galaxies and around E3 for 
the fainter ones.
The Hubble-type distribution of the faint ellipticals is
consistent with oblate symmetry, but the oblate hypothesis fails 
for the bright ellipticals.
However a distribution of triaxial intrinsic shapes can 
successfully reproduce the apparent shape data for either group.
The distribution of intrinsic, short-to-long axis ratios is peaked 
around
0.75 for bright galaxies and 0.65 for faint galaxies. 
Our results provide further evidence that elliptical galaxies 
should be divided into two, morphologically distinct families.

\end{abstract}

\section{Introduction}

In a previous paper (Tremblay \& Merritt 1995, hereafter Paper 
I) we presented a nonparametric technique for estimating the 
frequency function of elliptical galaxy intrinsic shapes.
We showed that the Djorgovski (1985) - Ryden (1992) axis ratio 
data for 171 elliptical galaxies was strongly inconsistent with the 
axisymmetric hypothesis but consistent with a number of triaxial
shape distributions.
The distribution of Hubble types was found to be broad or
bimodal, with weak maxima near E1 and E3.
Similar conclusions were reached by Fasano \& Vio (1991) and by
Ryden (1992, 1996), using the same and other data sets.

It has been clear for some time that faint elliptical 
galaxies are kinematically and morphologically distinct 
from brighter ellipticals, 
and this difference might be expected to manifest itself in the
respective distributions of axis ratios.
Elliptical galaxies brighter than $M_B\approx -20$ are generally 
slowly rotating, while fainter ellipticals exhibit roughly as much 
rotation as would be expected if their figures were 
centrifugally flattened (Davies {\it et al.} 1983).
Faint ellipticals also tend to exhibit disklike distortions and 
peaked, central density profiles, while bright ellipticals are 
more often ``boxy'' with shallow central profiles (Nieto {\it et al.} 1991;
Kormendy et al. 1995).
It is natural to assume that fainter ellipticals are 
oblate spheroids flattened by rotation,
while bright ellipticals might be triaxial, since in the absence 
of dynamically significant rotation there is no strong {\it a 
priori} case to be made for oblate symmetry (Binney 1978).

Here we extend the analysis of Paper I by investigating the 
dependence of the elliptical galaxy axis-ratio distribution 
on intrinsic luminosity.
We combine the Djorgovski - Ryden data set with Lauer \& 
Postman's (1994) sample of brightest cluster galaxies, whose 
intrinsic shape distribution was studied by Ryden, Lauer \& 
Postman (1993).
We find that the sample is effectively divided into two 
populations, with the division occurring near $M_B= -20$.
Brighter ellipticals have Hubble types that are narrowly 
clustered around E1-E2, while fainter ellipticals have Hubble 
types near E3.
Combining bright and faint galaxies into one sample produces the 
broad or double-peaked distribution seen in Paper I and in other
studies.
Furthermore, we find that while the distribution of Hubble types 
for the fainter ellipticals is consistent with oblate symmetry, 
the oblate hypothesis fails for the brighter ellipticals.
However a distribution of triaxial intrinsic shapes can be 
found that reproduces the apparent shape data for either subset
of galaxies.
Thus the frequency function of apparent axis ratios is consistent with -- 
though does not strictly imply -- a model in which fainter 
ellipticals are oblate and moderately flattened, while brighter 
ellipticals are rounder and triaxial.

\section{Data}

Our first set of galaxies is the sample introduced by Ryden (1992),
itself extracted from a program of CCD photometry of bright
ellipticals carried out by Djorgovski (1985).
As in Paper I, we continue to use the luminosity-weighted mean
axis ratios defined by Ryden (1992).

We obtained apparent magnitudes of the galaxies in the 
Djorgovski-Ryden sample from de Vaucouleurs {\it et al.} 
(1991, hereafter RC3).
We needed reliable, redshift-independent distances to each galaxy
in order to determine their absolute magnitudes -- defined throughout 
this study as total (asymptotic) blue magnitudes $M_B$.
The majority of the galaxies
in this sample have a distance derived from the
Dn-$\sigma$ relation (Dressler {\it et al.} 1987).
Our Dn-$\sigma$ distances came from the Mark II electronic
release catalogue of Burstein (1995) and from Faber {\it et al.} (1989).
A handful of the galaxies in the sample had their distances
measured using the surface brightness fluctuation (SBF) method 
(Tonry and Schneider 1988).
The tabulated Dn-$\sigma$ distances are given
in terms of corrected redshift velocities; we converted these 
redshifts to Mpc using a Hubble constant such that the resulting 
distance to the Virgo cluster is equal to 16 Mpc, equivalent to 
the SBF distance (Tonry, Ajhar \& Lupino 1990), and in agreement with
recent Cepheid data (Freedman {\it et al.} 1994; Pierce {\it et al.} 1994).
A total of 107 galaxies from the Djorgovski-Ryden sample had
redshift-independent distance estimates from one of these sources.

To this sample we added the 119 galaxies in the Lauer \& Postman 
(1994) sample of brightest cluster galaxies.
Luminosity-weighted axis ratios for these galaxies are given in 
Ryden, Lauer \& Postman (1993) and absolute magnitudes in Lauer 
\& Postman (1994).
Six galaxies were found to be in both samples, giving a total of 
220 galaxies in the combined set.

We carried out the analysis described below with both the 
combined sample, and with the Djorgovski-Ryden data alone.
Aside from the smaller degree of noise in the larger sample, we found no 
significant differences; hence we present results from only the 
combined sample below.

\section{Analysis}

We define $f(q,M_B)$ to be the joint distribution of 
elliptical galaxy apparent axis ratios $q$, $0<q\le 1$, 
and absolute magnitudes $M_B$.
Our goal is to construct an estimate of $f$, which we call
$\hat f$, from the data, then to operate on $\hat f$ 
to obtain estimates of the 
intrinsic shape distribution at any $M_B$.
The numerical inversion techniques for the last step are 
described in Paper I.

The estimation of multivariate density functions is the subject
of much current research in nonparametric statistics.
We used an ``adaptive product kernel'' estimator (Scott 1992, p. 149)
on the set of pairs $(q,M_B)$ to produce the estimate $\hat f(q,M_B)$. 
This estimator has the form
$$ \hat f(q,M_B)= {1 \over n h_q h_M} \sum_{i=1}^n \left[ 
l_i^{-2} K\left({q-q_i\over h_q l_i}\right) \times K
\left({M_B-{M_B}_i \over 
h_M l_i}\right)  \right] \eqno(1) 
$$
where $q_i$ and ${M_B}_i$ are the apparent axis ratio
and absolute magnitude of the $i^{th}$ galaxy. 
The function $K$ is called a kernel, and converts the discrete data
into a smooth continuous function.
We used the quartic kernel:
$$ K(x) = \cases {{15 \over 16} \left( 1 - x^2\right)^2 , 
&$-1 \leq x \leq 1$;\cr
0 , &otherwise.\cr}
$$
The quantities $h_q$ and $h_M$ are the window widths
of the respective variables, which must be chosen 
correctly if the estimate $\hat f$ is to lie close to the ``true'' 
function $f$. 
The quantity $l_i$ is a dimensionless variable which adapts the
window widths as a function of local density.
It is defined as
$$l_i = \left[ {\tilde f(q_i,M_{Bi}) \over g} \right]^{-1/2}
,$$ 
where $\tilde f(q_i,M_{Bi})$ is a ``pilot estimate'' of the 
density obtained using a fixed window width $(h_q,h_m)$, and $g$ 
is the geometric average of the $\tilde f(q_i,M_{Bi})$.
As is usually done, we reflected the data about the $q=1$ axis 
before constructing the estimates.

In Paper I we applied the ``unbiased cross-validation'' method 
(UCV) to determine
the optimal smoothing parameters in our one-dimensional study.
This method is readily adaptable to higher dimensions 
(e.g. Silverman 1986, p. 87).
The dependence of the UCV on the two smoothing parameters $h_q$ and
$h_M$ for our sample is shown in Figure 1. 
The smoothing parameters giving the minimum value of the UCV 
are $(h_q,h_M)=(0.11,0.8)$. 
The resulting, optimal estimate of the density $\hat f(q,M)$ is 
shown in Figure 2.
We have normalized $\hat f$ such that its integral along $q$ is unity at 
every fixed $M_B$ -- the distribution of absolute magnitudes is 
of no interest to us here.

Figure 2 shows two clear peaks: one near $q=0.85$ for the bright 
galaxies ($M_B\le -20$), and one near $q=0.7$ for the fainter galaxies.
Cuts of $\hat f(q,M_B)$ at three values of $M_B$ are shown in 
Figure 3.
The narrowness in $q$ of the bright-galaxy distribution is 
especially striking; the profile is well approximated as a 
Gaussian with a dispersion of only $0.08$ 
about the median value $q\approx 0.85$.
This dispersion is roughly equal to the adopted smoothing 
bandwidth in $q$ which implies that the real distribution may be 
even {\it more} strongly peaked in $q$ than Figure 3 suggests.

The result of Paper I -- a broad or bimodal distribution $f(q)$ 
-- can now be more completely understood: $f(q)$ consists of the 
superposition of two unimodal distributions, which 
have peaks at different values of $q$ for the bright and faint subsamples.

In Paper I we showed that the rapid falloff in $f(q)$ near $q=1$ 
was inconsistent with the axisymmetric hypothesis for galaxy 
intrinsic shapes.
Such a falloff is seen here (Figure 3) in the bright-galaxy subsample but 
not in the faint galaxies.
Estimates of the distribution of intrinsic axis ratios $\hat 
N(\beta)$ under the oblate and prolate hypotheses are shown in Figure 3 for 
three values of $M_B$; the dashed lines are 95\% bootstrap 
confidence bands.
(As discussed in Paper I, the estimates $\hat N$ are 
deconvolutions of $\hat f$ and should be constructed from 
estimates of $f$ that were obtained using larger values of the smoothing 
parameters than the ``optimal'' values derived above.
We chose $h_q=0.15$ when constructing $\hat f$ for use in 
computation of $\hat N$.)
The oblate and prolate hypotheses are both inconsistent at 
the 95\% level with the Hubble type distribution of the bright 
galaxies, but both are consistent with the faint galaxies.

As in Paper I, one might hope to successfully reproduce the 
Hubble-type distribution for the brighter ellipticals using a 
triaxial distribution of intrinsic shapes.
Figure 4 shows the result.
We have assumed that all galaxies are 
triaxial to the ``same'' degree, i.e. characterized by a fixed
value of $Z={1 -\beta_1 \over 1-\beta_2 }$, with $\beta_1$ and $\beta_2$
the two axis ratios, $1\geq\beta_1\geq\beta_2$.
Under such an assumption, 
one can indeed invert $\hat f(q)$ for 
the bright galaxies (as well as the fainter ones) without forcing 
$\hat N(\beta_2)$ to go negative near axis ratios of unity.

\section{Discussion}

Our primary result is that the distribution of Hubble types is 
significantly different for bright and faint elliptical galaxies, 
with the division occuring near $M_B= -20$.
Both families of galaxy exhibit unimodal distributions of apparent 
shapes, with the peak lying near E1.5 for 
bright ellipticals and near E3 for faint ellipticals.
The broader distributions seen in Paper I and in earlier studies may be 
interpreted as superpositions of these two, narrower frequency 
functions.

The distribution of Hubble types of the fainter ellipticals is 
consistent with the oblate, prolate and triaxial hypotheses; 
the intrinsic shape distribution inferred under 
any of these hypotheses has a peak in $c/a$ between $0.6$ and $0.7$.
The Hubble-type distribution of the bright ellipticals is inconsistent
with the axisymmetric hypothesis but is reproducable if one assumes
triaxiality; the intrinsic shape distribution is sharply peaked
at a short-to-long axis ratio of about $0.75$. 

Our results are consistent with earlier intrinsic-shape studies
that examined luminosity-selected groups of elliptical galaxies.
Ryden, Lauer and Postman's (1993) estimate of $f(q)$ for the subset 
of 119 brightest cluster galaxies included here 
looks quite similar to our estimate $\hat f(q,M_B)$ at $M_B=-20.5$, 
with a single peak near $q=0.83$ and a rapid falloff for $q$ near one.
Dwarf ellipticals (dE's), on the other hand, appear to be flatter 
on average than normal ellipticals (Ryden \& Terndrup 1994), 
similar to our result for low-luminosity E's.
However dwarf ellipticals differ in many fundamental ways from brighter,
``normal'' ellipticals and it is not clear that the two groups 
should be compared.
Fasano (1991) analyzed small samples of boxy and disky ellipticals
and found weak evidence for a difference in the apparent
ellipticity distributions; the difference was qualitatively similar to
what is seen here between bright and faint subsamples.

Our findings are consistent with a model in which fainter 
ellipticals are moderately flattened, oblate spheroids while bright 
ellipticals are more nearly round and triaxial.
However such a model is not compelled by our analysis since the 
Hubble-type distribution for the faint ellipticals is equally 
consistent with the triaxial hypothesis.
Nevertheless, the division of the {\it apparent} shape distribution 
into two groups at $M_B\approx -20$ is robust and implies a corresponding 
change in the distribution of intrinsic shapes at 
roughly this luminosity.

Our results contribute to the growing body of evidence that 
elliptical galaxies can be divided into two families that 
are morphologically and kinematically distinct (Bender 1988; Nieto,
Bender \& Surma 1991; Kormendy et al. 1995).
A number of plausible formation scenarios for these two families
are consistent with our results.
A greater role for dissipation in the formation of faint 
ellipticals would cause these galaxies to be both more highly 
flattened and more strongly rotating -- and possibly more oblate 
-- than bright ellipticals.
Bright ellipticals might form through the mergers of fainter
galaxies, a process that would likely make them rounder, more
slowly-rotating and possibly more triaxial than low-luminosity ellipticals.

Questions of formation aside, one can make inferences about 
intrinsic shapes based purely on the requirements of 
dynamical equilibrium.
Elliptical galaxies fainter than $M_B\approx -20$ have steep central
density cusps, $\rho\propto r^{-\gamma}$, $1\lesssim\gamma\lesssim 2$; 
bright ellipticals have shallower cusps, $0\lesssim \gamma\lesssim 1$ 
(Merritt \& Fridman 1995; Gebhardt et al. 1996).
Cusps steeper than $\gamma\approx 1$ will induce most of the boxlike
orbits in a triaxial galaxy to behave chaotically over astronomical
timescales (Merritt \& Valluri 1996).
One result is that triaxiality is difficult to maintain in a galaxy
with a steep cusp (Merritt \& Fridman 1996).
The change in the shape distribution seen here --- also
near $M_B=-20$ --- might reflect in part the influence of the 
central cusps on the global shapes.
However this mechanism would not naturally explain why the more
triaxial galaxies are less flattened.

The remarkably narrow distributions of axis ratios which we find 
at both high and low luminosities were quite unexpected.
It is important to understand what this result might imply about the
formation of elliptical galaxies.

\bigskip\bigskip
B. Ryden kindly supplied the luminosity-weighted axis ratios used
here and in Paper I, and prompted us to combine the Lauer-Postman 
sample with her own data.
D. Burstein sent us his Mark II library of estimated galaxy distances and
advised us in how to use it. 
This work was supported by NSF grant AST 90-16515 and NASA grant 
NAG 5-2803 to DM.

\clearpage

\figcaption[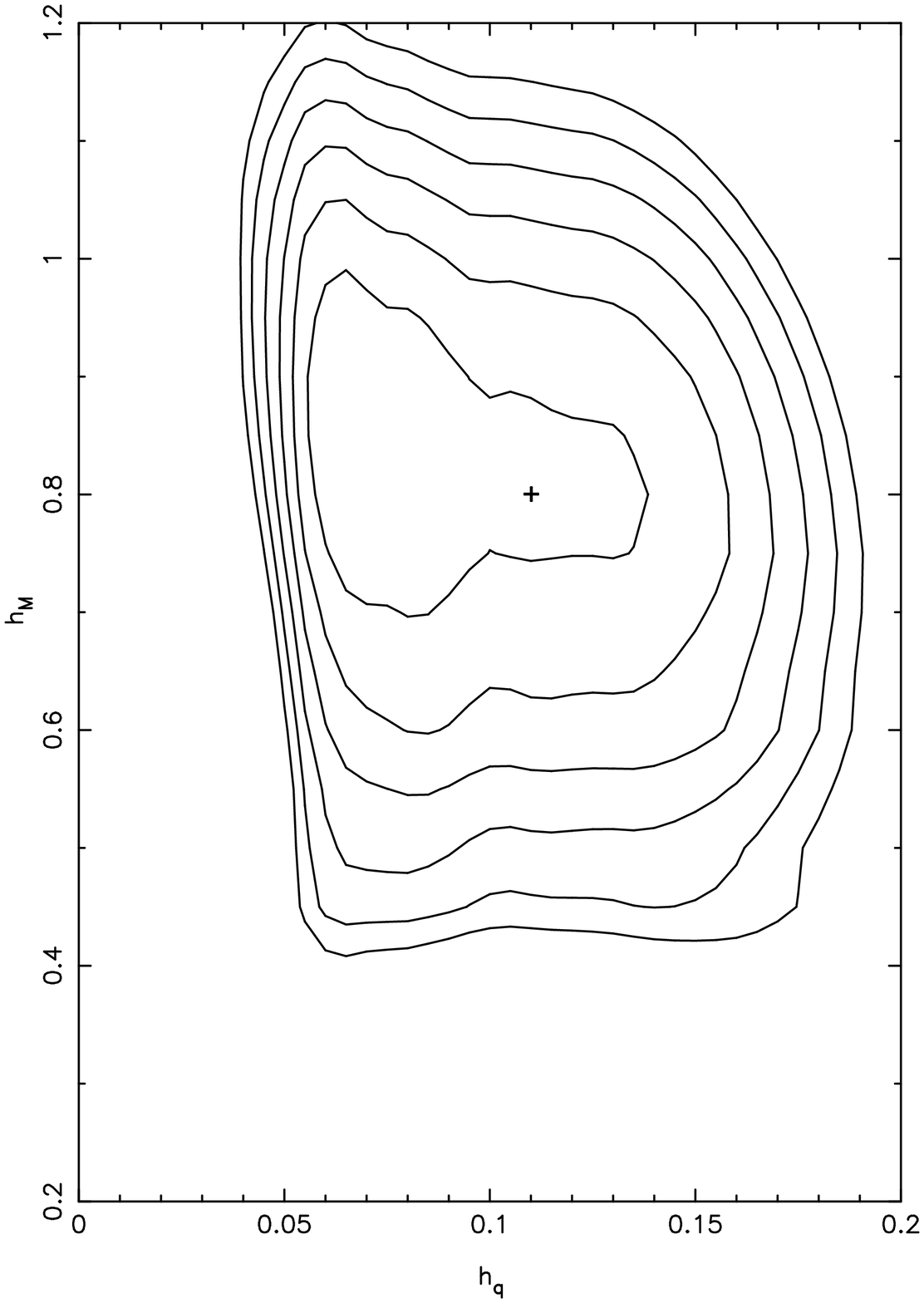]{\label{fig1}} Contours of UCV (see text) for 
various smoothing parameters. 
Contours are separated by 0.005; the cross indicates the position 
of the minimum.

\figcaption[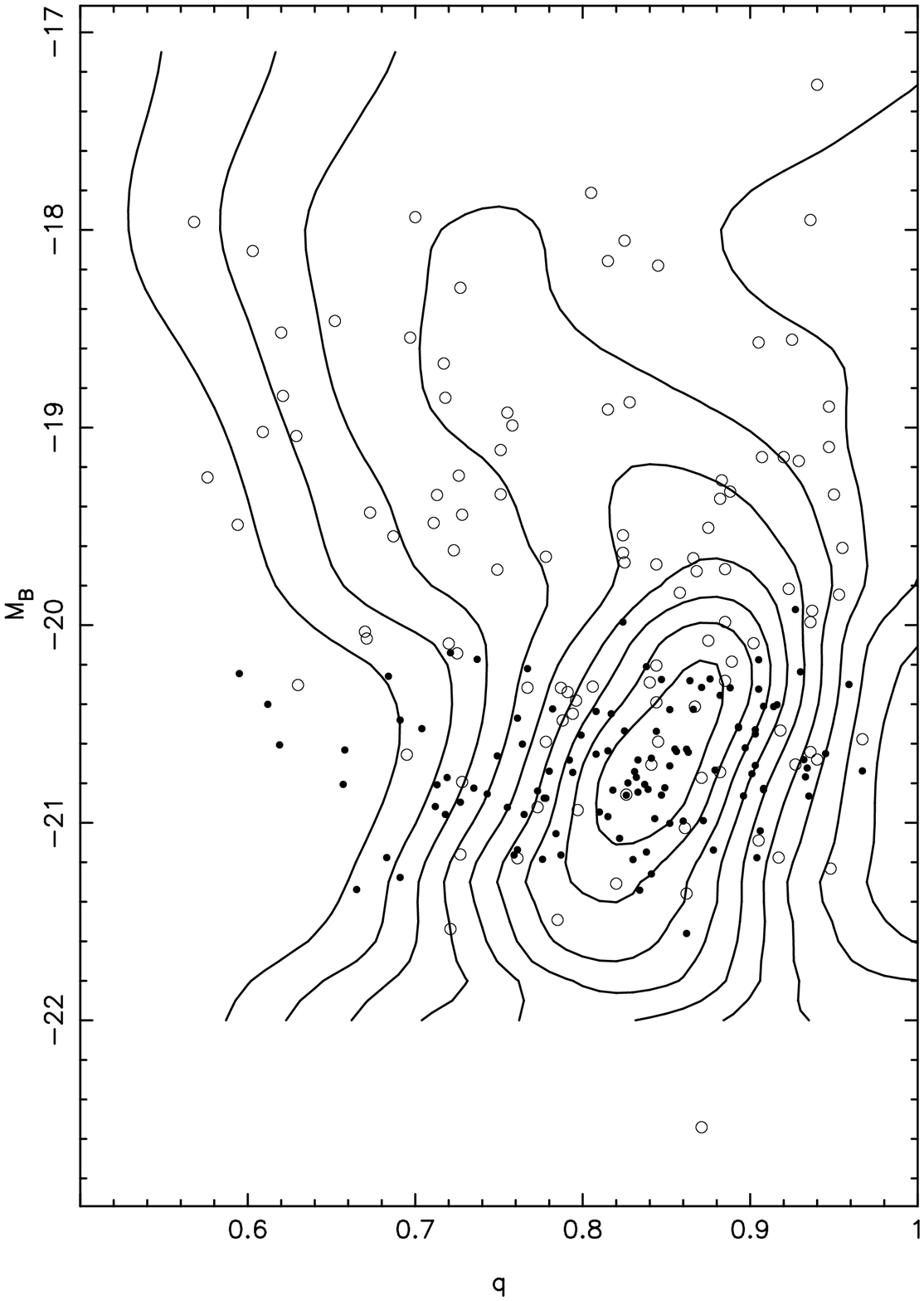]{\label{fig2}} Estimate of $f(q,M_B)$ for 
the 220 galaxies in our sample,
obtained via a product-kernel technique with different window widths
in the $q$ and $M_B$ directions.
The function $f$ has been normalized at every $M_B$ to give unit area
when integrated along $q$.
Circles are galaxies in the Djorgovski-Ryden sample; dots are from the
Lauer-Postman sample. 
The six galaxies in both samples are indicated by circles.
Contours are separated by 0.5 in $f$.

\figcaption[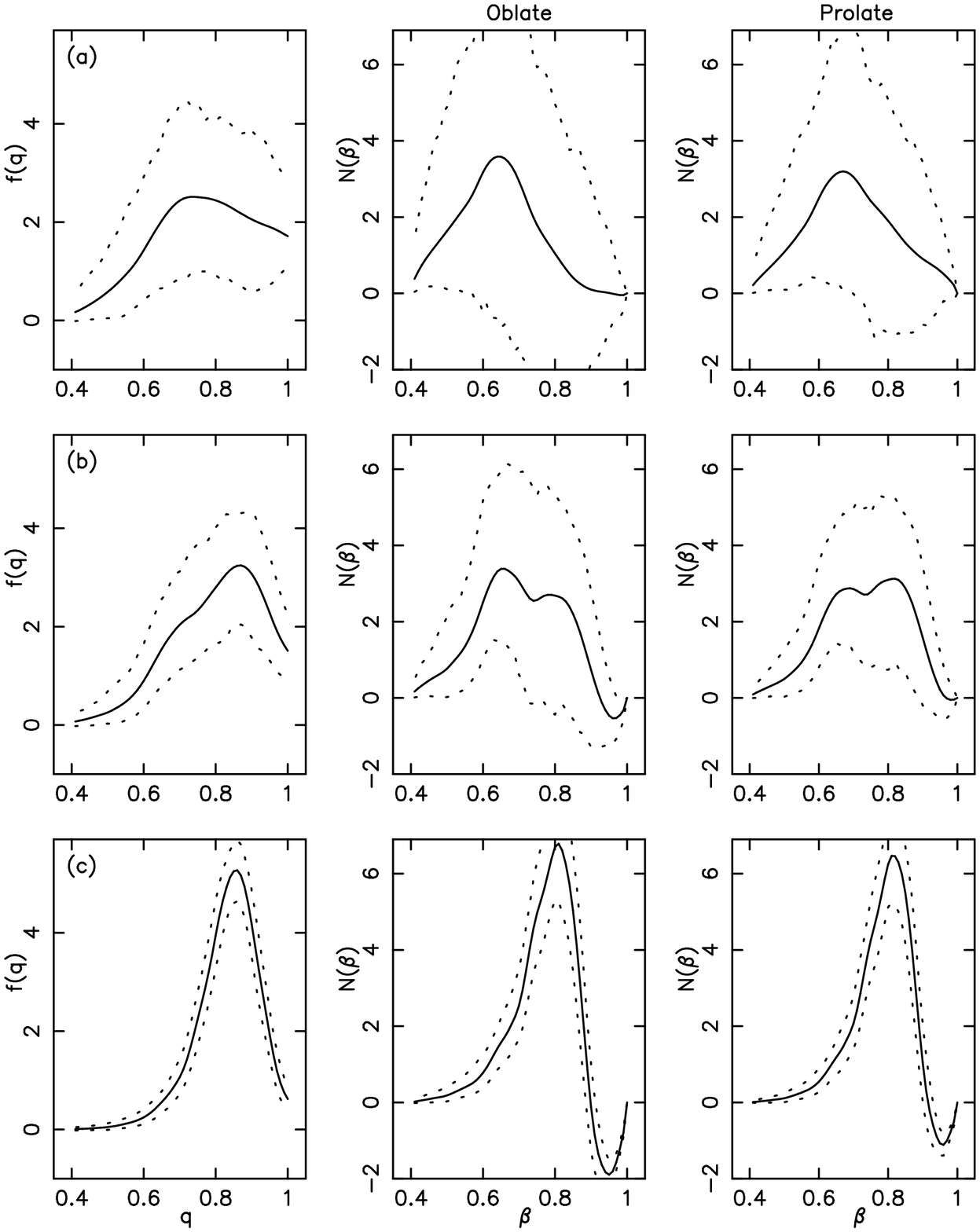]{\label{fig3}}
The dependence of the frequency function $\hat f(q,M_B)$ on $q$
at three values of $M_B$, and its oblate and prolate deconvolutions.
a) $M_B=-18.5$; b)$M_B=-19.5$; c) $M_B=-20.5$. 
Dashed lines are 95\% confidence bands on the estimates.

\figcaption[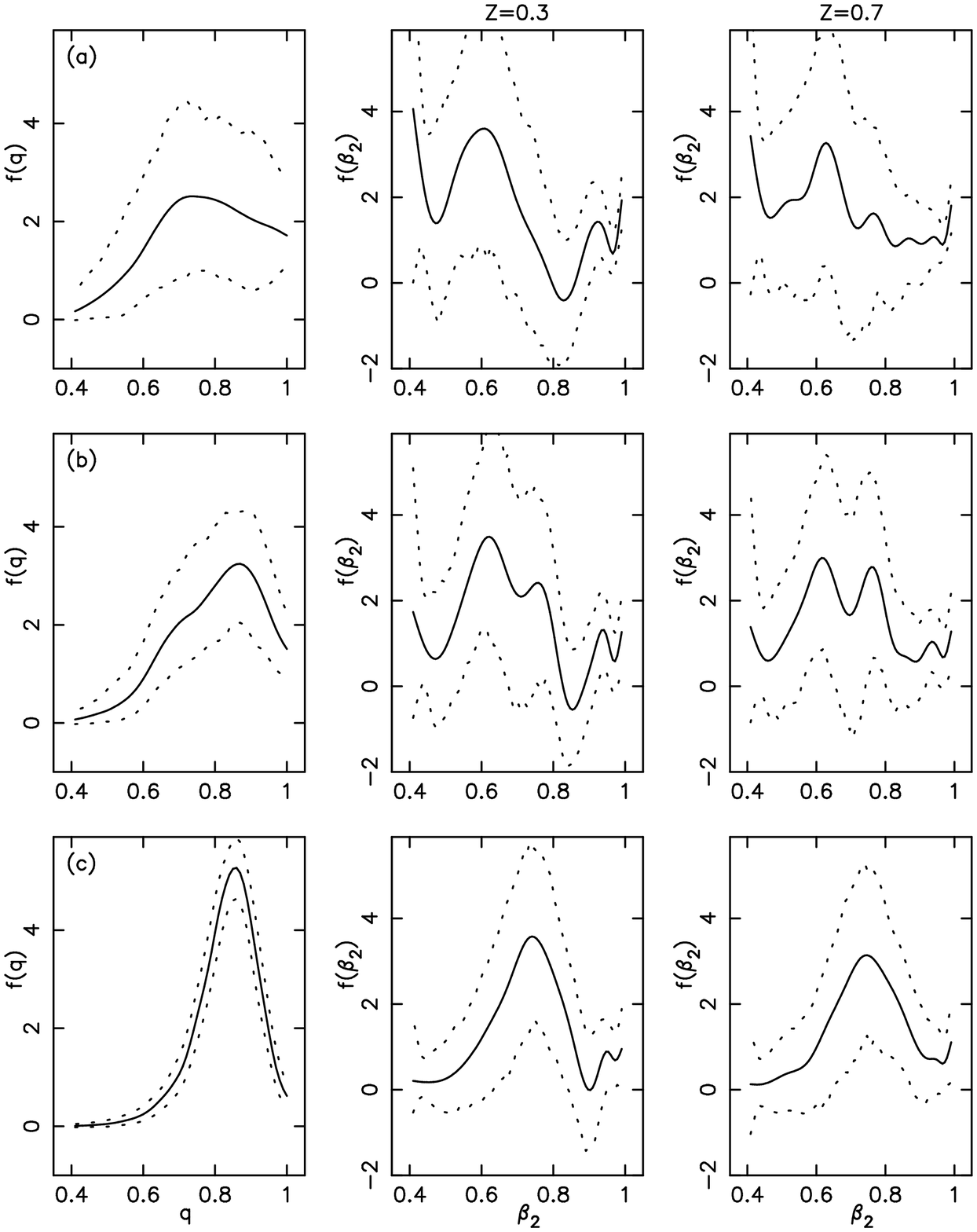]{\label{fig4}} The dependence of the
frequency function $\hat f(q,M_B)$ on $q$
at three values of $M_B$, and its deconvolution under the assumption
that all galaxies are triaxial to the same degree,
$Z = (1 - b/a)/(1- c/a) = $ constant.
a) $M_B=-18.5$; b)$M_B=-19.5$; c) $M_B=-20.5$. 
Dashed lines are 95\% confidence bands on the estimates.


\plotone{figure1.ps}

\plotone{figure2.ps}

\plotone{figure3.ps}

\plotone{figure4.ps}

\end{document}